# Decomposition Principles and Online Learning in Cross-Layer Optimization for Delay-Sensitive Applications

Fangwen Fu, Mihaela van der Schaar
Electrical Engineering Department, UCLA
{fwfu, mihaela}@ee.ucla.edu

*Abstract*— In this paper, we propose a general cross-layer optimization framework in which we explicitly consider both the heterogeneous and dynamically changing characteristics of delay-sensitive applications and the underlying time-varying network conditions. We consider both the independently decodable data units (DUs, e.g. packets) and the interdependent DUs whose dependencies are captured by a directed acyclic graph (DAG). We first formulate the cross-layer design as a non-linear constrained optimization problem by assuming complete knowledge of the application characteristics and the underlying network conditions. The constrained cross-layer optimization is decomposed into several cross-layer optimization subproblems for each DU and two master problems. These two master problems correspond to the resource *price* update implemented at the lower layer (e.g. physical layer, MAC layer) and the *impact factor* update for neighboring DUs implemented at the application layer, respectively. The proposed decomposition method determines the necessary message exchanges between layers for achieving the optimal cross-layer solution and it explicitly considers how the cross-layer strategies selected for one DU will impact its neighboring DUs as well as the DUs that depend on it. However, the attributes (e.g. distortion impact, delay deadline etc) of future DUs as well as the network conditions are often unknown in the considered real-time applications. The impact of current cross-layer actions on the future DUs can be characterized by a *state-value function* in the Markov decision process (MDP) framework. Based on the dynamic programming solution to the MDP, we develop a low-complexity cross-layer optimization algorithm using online learning for each DU transmission. This online optimization utilizes information only about the previous transmitted DUs and past experienced network conditions. This online algorithm can be implemented in real-time in order to cope with unknown source characteristics, network dynamics and resource constraints. Our numerical results demonstrate the efficiency of the proposed online algorithm.

# I. INTRODUCTION

To maximize its utility, a wireless user needs to jointly optimize the various protocol parameters and algorithms available at each layer of the OSI stack. This joint optimization of the transmission strategies at the various layers is referred to as *cross-layer optimization* [1][2].

## A. Related research

Cross-layer optimization has been extensively investigated in recent years in order to maximize the application's utility given the underlying time-varying and error-prone network characteristics. For instance, cross-layer optimization solutions for single-link communications [3][4][6], ad-hoc networks [7][8], and cellular networks [9] have been proposed. The majority of cross-layer optimization solutions can be divided into two main categories:

- *Static approaches*, in which the network conditions and application characteristics are described using static models (i.e. which remain unchanged over time), and the goal of the cross-layer optimization is to maximize a certain utility given such a static environment. Such solutions, including network utility maximization (NUM) [10] (and the references therein), do not explicitly consider and account for the time-varying source characteristics and network conditions, thereby resulting in suboptimal performance for the delay sensitive applications (e.g. wireless multimedia streaming) considered in this paper.

- *Sequential approaches*, in which the time-varying network conditions (e.g. channel conditions at the physical layer, allocated time/frequency bands at the MAC layer etc.) and application characteristics (e.g. packet arrivals, delay deadlines, distortion impact etc.) are explicitly modelled as (controlled) stochastic processes, and the goal is to sequentially determine the cross-layer actions over time to control this stochastic process such that the long-term utility is maximized [14][17]. The most important advantage of such sequential approaches is that they allow the wireless users to consider the experienced source and network dynamics (which are affected by both the uncertainty in the environment and the actions chosen by the wireless user) and, based on the users' knowledge about these dynamics up to that moment, select their cross-layer transmission strategies to maximize their utility *over time*. These solutions can significantly improve the transmission performance of delay-sensitive applications in time-varying wireless networks, as compared to the static approaches. However, current approaches consider simple models for both the time-varying application characteristics and dynamic network conditions which cannot satisfy the requirements of the delay-sensitive applications as explained below.

Based on the network dynamics and decision granularities in different layers, most sequential approaches for wireless transmission can be further classified into two categories: flow-based transmission decisions and DU-based transmission decisions. In the flow-based decision used in e.g. [3][4], the application data is assumed to be homogeneous (i.e. having the same distortion impact and same delay deadlines), and the network conditions are assumed to be time-varying (e.g. the network conditions are time-slotted and changes across the slots). The goal of the flow-based approaches is to optimize the "average" or "worst case" quality of service (QoS), e.g. average/worst case packet delay, packet loss rate, bit rate etc., for the supported applications. However, since the heterogeneous attributes of the packets in terms of delay deadlines and distortion impacts etc. are ignored, the flow-based approaches often result in suboptimal utilities for the delay-sensitive applications [24].

In DU-based transmission scenarios [11][15], each DU can contain one packet or multiple packets. Each DU is characterized by its distortion impact (e.g. the decrease in the application quality when that DU is lost), its packet length, the time at which the DU is ready for transmission and its delay deadline. For example, in video streaming applications, the DU can be one frame or one group of pictures, which may comprise multiple packets [11]. The decision is made for each DU to select the optimal transmission strategies across multiple layers such that the total quality of the application (e.g. the Peak Signal-to-Noise Ratio (PSNR) for multimedia streaming) is maximized. In [6], the optimal packet scheduling algorithm (i.e. DU-based) is developed for the transmission of a group of packets to minimize the consumed energy, while satisfying their common delay deadline. This optimal solution is obtained by assuming that the inter-arrival time and delay deadlines of the packets are known a priori. This solution also assumes that the underlying channel conditions are the same for all the packets. This packet scheduling algorithm is further extended to the case in which each packet has its own delay constraints in [5]. In [16], the authors further consider time-varying (time-slotted) channel conditions. However, the above papers do not consider the heterogeneity of the packets in terms of distortion impact on the supported applications (e.g. video streaming) etc. In [11], the video packets with various characteristics are scheduled considering a common delay deadline and an optimal solution (including optimal packet ordering and retransmission) is developed assuming that the underlying wireless channel is static. In [15], a DAG model is used to capture the media packet dependencies and, based on this, an optimal packet scheduling method is developed using dynamic programming [13]. However, the proposed solution disregards the dynamics and error protection capabilities at the lower layers (e.g. MAC and physical layers).

Summarizing, a general cross-layer optimization framework which simultaneously considers both the heterogeneous and dynamically changing DUs' attributes of delay-sensitive applications and the underlying time-varying network conditions is still missing. In this paper, we aim to develop a solution that addresses both of these challenges for the delay-sensitive applications such as multimedia transmission.

*B. Contribution of this paper*

We consider a DU-based approach, and assume that the cross-layer decisions are performed for each DU. We consider both the independently decodable DUs (i.e. they can be decoded independently without requiring the knowledge of other DUs) and the interdependent DUs (i.e. in order to be decoded, each DU requires those DUs it depends on to be decoded beforehand and these dependencies are expressed as a DAG). We first formulate a non-linear constrained optimization problem by assuming complete knowledge of the attributes[1] (including the time ready for transmission, delay deadlines, DU size and distortion impact and DAG-based dependencies) of the application DUs and the underlying network conditions. The formulations in [5][6][11][16] are special cases of the framework proposed in this paper.

---

[1] This is the case, for instance, when the multimedia data was pre-encoded and hinting files were created before transmission time [24]. However, in the real-time encoding case, these attributes are known just in time when the packets are deposited in the streaming buffer, which will be considered in Section V.

The constrained cross-layer optimization can be decomposed into several subproblems and two master problems as shown in Figure 1. We refer to each subproblem as Per-DU Cross-Layer Optimization (DUCLO) since it represents the cross-layer optimization for one DU. For the interdependent DUs, the DUCLOs are solved iteratively in a round-robin style. One master problem is called the Price Update (PU), which corresponds to the Lagrange multiplier (i.e. price of the resource) update associated with the considered resource constraint imposed at the lower layer (e.g. energy constraint); and the other master problem is called Neighboring Impact Factor Update (NIFU), which is implemented at the application layer. The NIFU corresponds to the update of the Lagrange multipliers (called Neighboring Impact Factors, NIFs) associated with the DU scheduling constraints between neighboring DUs[2]. It is clear that the decision granularity is one DU for DUCLO, two neighboring DUs for the NIFU, and all the DUs for the PU, as shown in Figure 1.

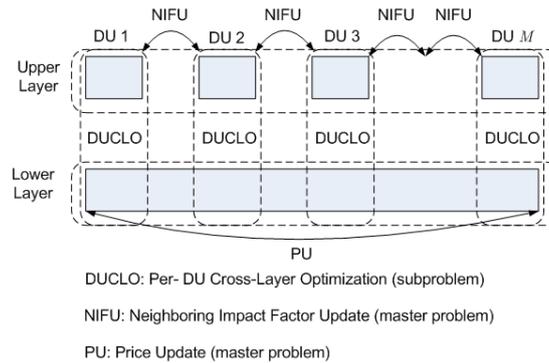

Figure 1. The decomposition of the cross-layer optimization and corresponding information update

The DUCLO problem for each DU is further separated into two optimizations: an optimization to determine the optimal scheduling time[3], which includes the time at which the transmission should start and it should be interrupted; and an optimization to determine the corresponding optimal transmission strategies at the lower layers (e.g. energy allocation at the physical layer, DU retransmission or FEC at the MAC layer). In this paper, we often refer to the application layer as the upper layer, while referring to the physical layer, MAC layer, network layer (or a combination of these layers) as the lower layer(s). As we will show in this paper, the proposed decomposition provides necessary message exchanges between layers and illustrates how the cross-layer strategies for one DU impact its neighboring DUs and the DUs it connects with in the DAG.

---

[2] These are consecutive packets generated by the source codec in the encoding/decoding order.
[3] The scheduling time is forwarded to the lower layer (e.g. the MAC layer) such that this layer can interrupt the transmission of the current packet and move to the next packet. A packet should be interrupted either because the DU's delay deadline has expired or because the next DU has higher precedence for transmission than the current DU due to its higher distortion impact.

In delay-sensitive real-time applications, the wireless user is often not allowed or cannot know the attributes of future DUs and corresponding network conditions. In other words, it only knows the attributes of previous DUs, and past experienced network conditions and transmission results. The message exchange mechanism developed based on the decomposition of the non-linear optimization is infeasible since it requires exact information about future DUs. However, when the distribution of the attributes and network conditions of DUs fulfil the Markov property [23], the cross-layer optimization can be reformulated as a MDP. Then the impact of the cross-layer action of the current DU on the future unknown DUs are characterized by a state-value function which quantifies the impact of the current DU's cross-layer action on the future DUs' distortion. Using the obtained decomposition principles developed for the online cross-layer optimization, we develop a low-complexity algorithm which only utilizes the available (causal) information to solve the online cross-layer optimization for each DU, update the resource price and learn the state-value function.

Thus, the difference between the methods proposed in this paper and those in [5][6][16][14] is that we explicitly take into account both the application characteristics and network dynamics, and determine decomposition principles for cross-layer optimization which adheres to the existing layered network architecture and illustrates the necessary massage exchanges between layers over time to achieve the optimal performance.

The rest of the paper is organized as follows. Section II formulates the cross-layer optimization problem for the independently decodable DUs as a non-linear constrained optimization assuming the knowledge of the characteristics of the supported application and underlying network conditions. Section III decomposes the optimization problem and presents the necessary message exchanges between layers and between neighboring DUs. Section IV further formulates the cross-layer optimization for interdependent DUs as a non-linear constrained optimization and presents the decomposed cross-layer optimization algorithm based on the decomposition principles developed in Section III. Section V presents an online cross-layer optimization for each DU transmission. Section VI shows some numerical results, followed by the conclusions in Section VII.

## II. PROBLEM FORMULATION

We assume that a wireless user streams delay-sensitive data over a time-varying wireless network. We focus on the DU-based cross-layer optimization. Specifically, the wireless user has $M$ DUs with individual delay constraints and different distortion impacts. In this section, we consider that the DUs are independently decodable and will discuss the cross-layer optimization for the interdependent DUs in Section IV. The time the DUs are ready for transmission is denoted by $t_i, i = 1, \cdots, M$. The delay deadline of each DU $i$ (i.e. the time before which the DUs must be received by the destination) is denoted by $d_i$, and thus, the following constraint needs to be satisfied: $d_i \geq t_i$. The DUs are transmitted in the First In First Out (FIFO) fashion (i.e. the same as the encoding/decoding order). The size of each DU $i$ is assumed to be $l_i$ bits. Each DU $i$ also has the distortion impact $q_i$ on the application. This distortion impact represents the decrease on the quality of the application when the entire DU is dropped [11][18]. Hence, each DU $i$ is associated with an attribute tuple $\psi_i = \{q_i, l_i, t_i, d_i\}$. In this section and the subsequent two sections, we assume that the attributes are known a priori for all DUs. In Section V, we will discuss the case in which the attributes of all the future DUs are unknown to the wireless user, as is the case in live encoding and transmission scenarios.

During the transmission, DU $i$ is delivered over the duration from time $x_i$ to time $y_i$ ($y_i \geq x_i$), where $x_i$ represents the starting transmission time (STX) and $y_i$ represents the ending transmission time (ETX). The choice of $x_i$ and $y_i$ represents the scheduling action of DU $i$, which is determined in the application layer. The scheduling action is denoted by $(x_i, y_i)$ satisfying the condition of $t_i \leq x_i \leq y_i \leq d_i$. At the lower layer (which can be one of the physical, MAC and network layers or combination of them), the wireless user experiences the average network condition $c_i \in \mathbb{R}_+$ during the transmission duration. For simplicity, we assume that the average network condition is independent of the scheduled time $(x_i, y_i)$, which can be the case when the network condition is slowly changing. The wireless user can deploy the transmission action $a_i \in \mathcal{A}$ based on the experienced network condition. The set $\mathcal{A}$ represents the possible transmission actions that the wireless user can choose. The transmission action at the lower layer can be, for example, the number of DU transmission retry (e.g. ARQ) at the MAC layer, and energy allocation at the physical layer.

When the wireless user deploys the transmission action $a_i$ under the network condition $c_i$, the expected distortion of DU $i$ due to the imperfect transmission in the network is represented by $Q_i(x_i, y_i, a_i) = q_i p_i(x_i, y_i, a_i)$[4], where $p_i(x_i, y_i, a_i)$ can be the probability that DU $i$ is lost as in [15] or the distortion decaying function[5] due to partial data of DU $i$ being received as in [18]. The resource cost incurred by its transmission is represented by $w_i(x_i, y_i, a_i) \in \mathbb{R}_+$. In addition, we assume that the functions $p_i(x_i, y_i, a_i)$ and $w_i(x_i, y_i, a_i)$ satisfy the following conditions:

C1 (Monotonicity): $p_i(x_i, y_i, a_i)$ is a non-increasing function of the difference $y_i - x_i$ and the transmission action $a_i$.

C2 (Convexity): $p_i(x_i, y_i, a_i)$ and $w_i(x_i, y_i, a_i)$ are convex functions of the difference $y_i - x_i$ and the transmission action $a_i$.

Condition C1 means that the expected distortion will be reduced by increasing the difference $y_i - x_i$, since this results in a longer transmission time which increases the chance DU $i$ will be successfully transmitted. In condition C2, the convexities of $p_i$ and $w_i$ are assumed to simplify the analysis. This assumption is satisfied in most scenarios, as will be shown in Section VI.

Based on the description above, the cross-layer optimization for the delay-sensitive application over the wireless network is to find the optimal scheduling action (i.e. determining the STX $x_i$ and ETX $y_i$ for each DU) at the application layer and, under the scheduled time, the optimal transmission action $a_i$ at the lower layer. The goal of the cross-layer optimization is to minimize the expected average distortion experienced by the delay-sensitive application. This cross-layer optimization may also be constrained on the available resources at the lower layer (e.g. energy at the physical layer). Then, the cross-layer optimization problem with complete knowledge (referred to as CK-CLO) can be formulated as:

$$\min_{x_i, y_i, a_i, i=1,\cdots,M} \frac{1}{M} \sum_{i=1}^{M} Q_i(x_i, y_i, a_i)$$
$$\text{s.t. } x_i \leq y_i, x_i \geq t_i, y_i \leq d_i, x_{i+1} \geq y_i, a_i \in \mathcal{A}, \quad \text{(CK-CLO)}$$
$$\frac{1}{M} \sum_{i=1}^{M} w_i(x_i, y_i, a_i) \leq W.$$

where the constraint $x_{i+1} \geq y_i$ indicates that DU $i+1$ has to be transmitted after DU $i$ is transmitted (i.e. FIFO), and the last line in the CK-CLO problem indicates the resource constraint in which $W$ is the average resource budget (e.g. the available energy for transmission).

---

[4] We consider here that the distortion of the independently decodable DUs is not affected by other DUs, as in [20].

[5] The distortion decaying function represents the fraction of the distortion remained after the (partial) data are successfully transmitted. For example, when the source is encoded in a scalable way, the distortion function is given by $D = Ke^{-\theta R}$ when $R$ bits has been received [18]. In this case, the distortion decaying function is given as $p_i(x_i, y_i, a_i) = e^{-\theta_i R_i(x_i, y_i, a_i)}$ and $q_i = K$.

## III. DECOMPOSITION FOR CROSS-LAYER OPTIMIZATION

In this section, we discuss how the cross-layer optimization in the CK-CLO problem can be decomposed using duality theory [12], what information has to be updated among DUs at each layer and what messages have to be exchanged across multiple layers. Such decomposition principles are important for developing optimal cross-layer solutions, because it adheres to the current layered network architecture.

*A. Lagrange dual problem*

We first relax the constraints in the CK-CLO problem by introducing the Lagrange multiplier $\lambda \geq 0$ associated with the resource constraint and Lagrange multiplier vector $\boldsymbol{\mu} = [\mu_1, \cdots, \mu_{M-1}]^T \geq \mathbf{0}$, whose elements are associated with the constraint $x_{i+1} \geq y_i, \forall i$. The corresponding Lagrange function is given as

$$L(\boldsymbol{x}, \boldsymbol{y}, \boldsymbol{a}, \lambda, \boldsymbol{\mu}) = \frac{1}{M} \sum_{i=1}^{M} Q_i(x_i, y_i, a_i) + \lambda \left( \frac{1}{M} \sum_{i=1}^{M} w_i(x_i, y_i, a_i) - W \right) + \sum_{i=1}^{M-1} \mu_i (y_i - x_{i+1}), \quad (1)$$

where $\boldsymbol{x} = [x_1, \cdots, x_M]$, $\boldsymbol{y} = [y_1, \cdots, y_M]$ and $\boldsymbol{a} = [a_1, \cdots, a_M]$.

Then, the Lagrange dual function is given by

$$g(\lambda, \boldsymbol{\mu}) = \min_{\substack{x_i, y_i, a_i, \\ i=1,\cdots,M}} \left\{ \frac{1}{M} \sum_{i=1}^{M} Q_i(x_i, y_i, a_i) + \lambda \left( \frac{1}{M} \sum_{i=1}^{M} w_i(x_i, y_i, a_i) - W \right) + \sum_{i=1}^{M-1} \mu_i (y_i - x_{i+1}) \right\} \quad (2)$$
$$\text{s.t. } x_i \leq y_i, x_i \geq t_i, y_i \leq d_i, a_i \in \mathcal{A}, i=1,\cdots,M$$

The dual problem (referred to as CK-DCLO) is then given by

$$\max_{\lambda \geq 0, \boldsymbol{\mu} \geq 0} g(\lambda, \boldsymbol{\mu}) \quad \text{(CK-DCLO)}$$

where $\boldsymbol{\mu} \geq 0$ denotes the component-wise inequality. The CK-DCLO dual problem can be solved using the subgradient method as shown next.

The subgradients of the dual function are given by $h_\lambda = \left( \frac{1}{M} \sum_{i=1}^{M} w_i(x_i, y_i, a_i) - W \right)$ with respect to the variable $\lambda$ and $h_{\mu_i} = (y_i - x_{i+1})$ with respect to the variable $\mu_i$ [12]. The CK-DCLO problem can then be iteratively solved using the subgradients to update the Lagrange multipliers as follows.

*Price-Updating*:

$$\lambda^{k+1} = \left( \lambda^k + \alpha^k \left( \frac{1}{M} \sum_{i=1}^{M} w_i(x_i, y_i, a_i) - W \right) \right)^+ \tag{3}$$

and *NIF Updating*:

$$\mu_i^{k+1} = \left( \mu_i^k + \beta_i^k (y_i - x_{i+1}) \right)^+, \tag{4}$$

where $z^+ = \max\{z, 0\}$ and $\alpha_k$ and $\beta_i^k$ are the update step size and satisfy the following conditions: $\sum_{k=1}^{\infty} \alpha^k = \infty, \sum_{k=1}^{\infty} (\alpha^k)^2 < \infty$ and $\sum_{k=1}^{\infty} \beta_i^k = \infty, \sum_{k=1}^{\infty} (\beta_i^k)^2 < \infty$ [6]. The proof of convergence is given in [12].

From the subgradient method, we note that the Lagrange multiplier $\lambda$ is updated based on the consumed resource and available budget, which is interpreted as the "price" of the resource and it is determined at the lower layer, while the Lagrange multiplier vector $\boldsymbol{\mu}$ is updated based on the scheduling time of the neighboring DUs, which is interpreted as the neighboring impact factors and is determined at the application layer. The update is also illustrated in Figure 2, and the details of this figure are presented subsequently. Since the CK-CLO problem is a convex optimization, the duality gap between the CK-CLO and CK-DCLO problems is zero, which is further demonstrated in Section VI. Based on the multiplier update given in Eqs. (3) and (4), we can make the following remark, which is essential for implementing practical cross-layer designs.

**Remark 1**: The update of the Lagrange multipliers $\lambda$ and $\boldsymbol{\mu}$ can be performed separately in the different layers, thereby automatically adhering to the layered network architecture.

---

[6] These conditions are required to enforce the convergence of the subgradient method. The choice of $\alpha^k$ and $\beta_i^k$ trades off the speed of convergence and performance obtained. One example is $\alpha^k = \beta_i^k = 1/k$.

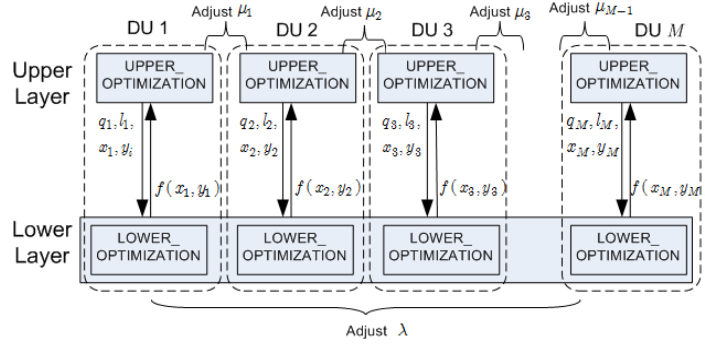

Figure 2. Message exchange between layers and between neighboring DUs

### B. Decomposition for Lagrange dual function

Given the Lagrange multipliers $\lambda$ and $\mu$, the dual function shown in Eq. (2) is separable and can be decomposed into $M$ DUCLO problems:

DUCLO problem $i \in \{1,\cdots,M\}$:

$$\min_{x_i,y_i,a_i} \frac{1}{M}Q_i(x_i,y_i,a_i) + \frac{\lambda}{M}w_i(x_i,y_i,a_i) - \mu_{i-1}x_i + \mu_i y_i \quad (5)$$
$$s.t. \ x_i \leq y_i, x_i \geq t_i, y_i \leq d_i, a_i \in \mathcal{A}$$

where $\mu_0 = 0$ and $\mu_M = 0$. Given the Lagrange multipliers $\lambda$ and $\mu$, each DUCLO problem is independently optimized. From Eq. (5), we note that all the DUCLO problems share the same Lagrange multiplier $\lambda$, since the budget constraint at the lower layer is imposed on all the DUs (see Figure 2). We also note that each DUCLO problem $i$ shares the same Lagrange multiplier $\mu_{i-1}$ with DUCLO problem $i-1$ and $\mu_i$ with DUCLO problem $i+1$ (see Figure 2). Compared to the traditional myopic algorithm in which each DU is transmitted greedily without considering its impact on future DUs (e.g. flow-based approaches), the DUCLO problems presented here automatically take into account the impact of the scheduling for the current DU on its neighbours.

***Remark 2***: The impact between the independently decodable DUs takes place only through the Lagrange multipliers $\lambda$ and $\mu$. Hence, we can separately find the cross-layer actions for each DU by estimating the Lagrange multipliers $\lambda$ and $\mu$, which will be used in the online implementation discussed in Section V.

*C.    Layered Solution to the DUCLO problem*

In this section, we describe how the DUCLO problem can be separated into two layered subproblems and what messages should be exchanged between layers. Given the Lagrange multipliers $\lambda$ and $\mu$, the DUCLO in Eq. (5) can be rewritten as

$$\min_{x_i,y_i}\left\{\min_{a_i\in\mathcal{A}}\left\{\frac{1}{M}Q_i(x_i,y_i,a_i)+\frac{\lambda}{M}w_i(x_i,y_i,a_i)\right\}-\mu_{i-1}x_i+\mu_i y_i\right\} \quad (6)$$
$$s.t. \quad x_i \leq y_i, x_i \geq t_i, y_i \leq d_i,$$

The inner optimization in Eq. (6) is performed at the lower layer and aims to find the optimal transmission action $a_i^*$, given STX $x_i$ and ETX $y_i$. This optimization is referred to as *LOWER_OPTIMIZATION*:

$$f(x_i,y_i)=\min_{a_i\in\mathcal{A}}\frac{1}{M}Q_i(x_i,y_i,a_i)+\frac{\lambda}{M}w_i(x_i,y_i,a_i) \quad (7)$$

The LOWER_OPTIMIZATION requires the information of the scheduling time $(x_i,y_i)$, distortion impact $q_i$ and DU size $l_i$ which are obtained from the upper layer and the information of transmission actions $a_i$ and price of resource $\lambda$, which are obtained at the lower layer.

The outer optimization in Eq. (6) is performed at the upper layer and aims to find the optimal STX $x_i$ and ETX $y_i$, given the solution to the lower optimization in Eq. (7). This optimization is referred to as the *UPPER_OPTIMIZATION*:

$$\min_{x_i,y_i} f(x_i,y_i)-\mu_{i-1}x_i+\mu_i y_i$$
$$s.t. \quad x_i \leq y_i, x_i \geq t_i, y_i \leq d_i, \quad (8)$$

The UPPER_OPTIMIZATION requires the information of $f(x_i,y_i)$, which can be interpreted as the best response to $(x_i,y_i)$ performed at the lower layer, and information of $\mu_{i-1}$ and $\mu_i$ which are obtained at the upper layer.

Hence, given the message $\{q_i,l_i,x_i,y_i\}$, the LOWER_OPTIMIZATION can optimally provide $a_i^*$ and the best response function $f(x_i,y_i)$. Given the function $f(x_i,y_i)$, the UPPER_OPTIMIZATION tries to find the optimal STX $x_i^*$ and ETX $y_i^*$. This message exchange is illustrated in Figure 2.

Since $Q_i(x_i,y_i,a_i)$ and $w_i(x_i,y_i,a_i)$ are convex functions of the difference $y_i-x_i$ and $a_i$, the LOWER_OPTIMIZATION and UPPER_OPTIMIZATION are both convex optimization problems and can be efficiently solved using well-known convex optimization algorithms such as the interior-point methods [21].

***Remark 3***: This layered solution for one DU provides the necessary message exchanges between the upper layer and lower layer, and illustrates the role of each layer in the cross-layer optimization. Specifically, the application layer works as a "guide" which determines the optimal STX and ETX by taking into account the best response $f(x_i, y_i)$ of the lower layer, while the lower layer works as a "follower", which only needs to determine the best response $f(x_i, y_i)$, given the scheduling time $(x_i, y_i)$ determined by the upper layer.

In summary, the algorithm for solving the CK-CLO problem is illustrated in Algorithm 1.

Algorithm 1: Algorithm for solving the CK-CLO problem

**Initialize** $\lambda^0, \boldsymbol{\mu}^0, \lambda^1, \boldsymbol{\mu}^1, \varepsilon, k = 1$
**While** ($\left|\lambda^k - \lambda^{k-1}\right| + \left\|\boldsymbol{\mu}^k - \boldsymbol{\mu}^{k-1}\right\| > \varepsilon$ or $k = 1$)
   **For** $i = 1, \cdots, M$
     Layered solution to DUCLO for DU $i$
  **End**
  Compute $\lambda^{k+1}, \boldsymbol{\mu}^{k+1}$ as in Eqs. (3) and (4).
  $k \leftarrow k + 1$
**End**

## IV. CROSS-LAYER OPTIMIZATION FOR INTERDEPENDENT DUS

In this section, we consider the cross-layer optimization for interdependent DUs. The interdependencies can be expressed using a DAG. One example for video frames is given in Figure 3. (More examples can be found in [15]). Each node of the graph represents one DU and each edge of the graph directed from DU $i$ to DU $i'$ represents the dependence of DU $i$ on DU $i'$. This dependency means that the distortion impact of DU $i$ depends on the amount of successfully received data in DU $i'$. We can further define the partial relationship between two DUs which may not be directly connected, for which we write $i' \prec i$ if DU $i'$ is an ancestor of DU $i$ or equivalently DU $i$ is a descendant of DU $i'$ in the DAG. The relationship $i' \prec i$ means that the distortion (or error) is propagated from DU $i'$ to DU $i$.

The error propagation function from DU $i'$ to DU $i$ is represented by $e_{i'}(x_{i'}, y_{i'}, a_{i'}) \in [0,1]$ [7] which is assumed to be a decreasing convex function of the difference $y_{i'} - x_{i'}$ and $a_{i'}$. Then, the distortion impact of DU $i$ can be computed as

$$Q_i(x_i, y_i, a_i) = q_i - q_i \left[ (1 - p_i(x_i, y_i, a_i)) \prod_{k \prec i}(1 - e_k(x_k, y_k, a_k)) \right]. \quad (9)$$

If DU $i$ cannot be decoded because one of its ancestor is not successfully received and $p_i(x_i, y_i, a_i)$ represents the loss probability of DU $i$, then $e_i(x_i, y_i, a_i) = p_i(x_i, y_i, a_i)$ as in [15].

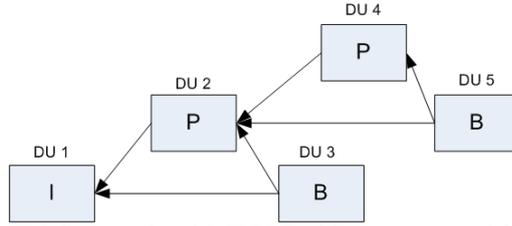

Figure 3. DAG example with IBPBP video compressed frames

The primary problem of the cross-layer optimization for the interdependent DUs is the same as in the CK-CLO problem by replacing $Q_i(x_i, y_i, a_i)$ with the formula in Eq. (9). The difference from the CK-CLO problem is that $Q_i(x_i, y_i, a_i)$ here depends on the cross-layer actions of its ancestors and $Q_i(x_i, y_i, a_i)$ may not be a convex function of all the cross-layer actions $(x_k, y_k, a_k) \forall k \preceq i$, although $e_k(x_k, y_k, a_k)$ is a convex function of $(x_k, y_k, a_k)$. However, we note that, given $(x_k, y_k, a_k) \forall k \prec i$, $Q_i(x_i, y_i, a_i)$ is a convex function of $(x_i, y_i, a_i)$. We will use this property to develop a dual solution for the original non-convex problem and we will quantify the duality gap in the simulation section.

The derivative of the dual problem is the same as the one in Section III. By replacing $Q_i(x_i, y_i, a_i)$ with the formula in Eq. (9), the Lagrange dual function shown in Eq. (2) becomes

$$g(\lambda, \boldsymbol{\mu}) = \min_{\substack{x_i, y_i, a_i, \\ i=1,\cdots,M}} \left\{ \frac{1}{M} \sum_{i=1}^{M} \left( q_i - q_i(1 - p_i(x_i, y_i, a_i)) \prod_{k \prec i}(1 - e_k(x_k, y_k, a_k)) \right) + \lambda \left( \frac{1}{M} \sum_{i=1}^{M} w_i(x_i, y_i, a_i) - W \right) + \sum_{i=1}^{M-1} \mu_i(y_i - x_{i+1}) \right\}. \quad (10)$$

$$\text{s.t.} \quad x_i \leq y_i, x_i \geq t_i, y_i \leq d_i, a_i \in \mathcal{A}, i = 1, \cdots, M$$

---

[7] In general, the error propagation function $e_{i'}(x_{i'}, y_{i'}, a_{i'})$ of DU $i'$ also depends on which DU it will affect [20]. For simplicity, we assume the error propagation function only depends on the current DU and does not depend on the DU it will affect. In this paper, to simplify the analysis, we do not consider the impact of error concealment strategies. Such strategies could be used in practice, and this will not affect the proposed methodology for cross-layer optimization.

Due to the interdependency, this dual function cannot be simply decomposed into the independent DUCLO problems as shown in Eq. (5). However, the dual function can be computed DU by DU assuming the cross-layer actions of other DUs is given, as shown in [15]. Specifically, given the Lagrange multipliers $\lambda, \mu$, the objective function in Eq. (10) is denoted as $G((x_1,y_1,a_1),\cdots,(x_M,y_M,a_M),\lambda,\mu)$. When the cross-layer actions of all DUs except DU $i$ are fixed, the DUCLO for DU $i$ is given by

$$\min_{x_i \leq y_i, x_i \geq t_i, y_i \leq d_i, a_i \in \mathcal{A}} G((x_1,y_1,a_1),\cdots,(x_i,y_i,a_i),\cdots,(x_M,y_M,a_M),\lambda,\mu)$$
$$= \min_{x_i \leq y_i, x_i \geq t_i, y_i \leq d_i, a_i \in \mathcal{A}} \left( \frac{1}{M} Q'_i(x_i,y_i,a_i) + \frac{\lambda}{M} w_i(x_i,y_i,a_i) - \mu_{i-1} x_i + \mu_i y_i \right) + \theta_i \quad (11)$$

where
$Q'_i(x_i,y_i,a_i) =$

$$\frac{1}{M} q_i p_i(x_i,y_i,a_i) \prod_{k \prec i} e_k(x_k,y_k,a_k) - (1 - e_i(x_i,y_i,a_i)) \left( \sum_{i' \succ i} q_{i'}(1 - p_{i'}(x_{i'},y_{i'},a_{i'})) \prod_{\substack{k \prec i' \\ k \neq i}} (1 - e_k(x_k,y_k,a_k)) \right), \quad (12)$$

and $\theta_i$ represents the remaining part in Eq. (10), which does not depend on the cross-layer action $(x_i,y_i,a_i)$. It is easy to show that the optimization over the cross-layer action of DU $i$ in Eq. (11) is a convex optimization, which can be solved in a layered fashion as shown in Section III.C.

As discussed in [15], $Q'_i(x_i,y_i,a_i)$ can be interpreted as the sensitivity to (or impact of) the imperfect transmission of DU $i$, i.e. the amount by which the expected distortion will increase if the data of DU $i$ is fully received, given the cross-layer actions of other DUs. It is clear that the DUCLO for DU $i$ is solved only by fixing the cross-layer actions of other DUs, unlike the solutions for the independently decodable DUs which do not require the knowledge of other DUs.

Then, the optimization in Eq. (10) can be solved using the block coordinate descent method [12], as described next. Given the current optimizer $((x_1^n,y_1^n,a_1^n),\cdots,(x_M^n,y_M^n,a_M^n))$ at iteration $n$, the optimizer at iteration $n+1$, $((x_1^{n+1},y_1^{n+1},a_1^{n+1}),\cdots,(x_M^{n+1},y_M^{n+1},a_M^{n+1}))$ is generated according to the iteration

$$\begin{aligned}(x_i^{n+1},y_i^{n+1},a_i^{n+1}) = \arg \min_{x_i \leq y_i, x_i \geq t_i, y_i \leq d_i, a_i \in \mathcal{A}} \\ G\big((x_1^{n+1},y_1^{n+1},a_1^{n+1}),\cdots,(x_{i-1}^{n+1},y_{i-1}^{n+1},a_{i-1}^{n+1}),(x_i,y_i,a_i),(x_{i+1}^n,y_{i+1}^n,a_{i+1}^n),\cdots,(x_M^n,y_M^n,a_M^n),\lambda,\mu\big)\end{aligned} \quad (13)$$

At each iteration, the objective function is decreased compared to that of the previous iteration and the objective function is lower bounded (greater than zero). Hence, this block coordinate descent method converges to the locally optimal solution to the optimization in Eq. (10), given the Lagrange multipliers $\lambda$ and $\mu$.

***Remark 4***: From Eq. (11), we note that, when we focus on the cross-layer optimization for DU $i$, besides the resource price $\lambda$ and NIF $\mu_{i-1}$ and $\mu_i$ as requested for the independently decodable DU, we further need some additional information: the interdependencies with other DUs (expressed by the DAG) and the values of $p_k(x_k, y_k, a_k)$ and $e_k(x_k, y_k, a_k)$ of all DUs $k$ connected with DU $i$. For real-time applications, the information of future DUs is often unavailable when DU $i$ is transmitted. We show in Section V how this information can be estimated online.

## V.  ONLINE CROSS-LAYER OPTIMIZATION WITH INCOMPLETE KNOWLEDGE

The cross-layer optimization formulated in Sections II and IV assumes complete a-priori knowledge of the DUs' attributes and the network conditions. However, in real-time applications, this knowledge is only available just before the DUs are transmitted. Furthermore, the cross-layer optimization algorithms based on the decomposition principles presented in Sections III and IV require multiple iterations (as shown in Sections VI.B and VI.C) to converge, which may be difficult to implement for real-time applications. To deal with the real-time transmission scenario, we propose a low-complexity online cross-layer optimization algorithm motivated by the decomposition principles developed in Sections III and IV.

### A. *Online optimization using learning for independent DUs*

In this section, we assume that the DUs can be independently decoded and that the attributes and network conditions dynamically change over time. The random versions of the time the DU is ready for transmission, delay deadline, distortion impact and network condition are denoted by $T_i, D_i, L_i, \mathcal{Q}_i, C_i$, respectively. We assume that both the inter-arrival interval (i.e. $T_{i+1} - T_i$) and the life time (i.e. $D_i - T_i$) of the DUs are i.i.d. The other attributes of each DU and the experienced network condition are also i.i.d. random variables independent of other DUs. We further assume that the user has an infinite number of DUs to transmit. Then, the cross-layer optimization with complete knowledge presented in the CK-CLO problem becomes a cross-layer optimization with incomplete knowledge (referred to as ICK-CLO) as shown below:

$$\min_{x_i,y_i,a_i,\forall i} \lim_{N\to\infty} \frac{1}{N}\sum_{i=1}^{N} \mathop{E}_{T_i,D_i,L_i,\mathcal{Q},C_i} Q_i(x_i,y_i,a_i)$$
$$s.t.\ x_i \geq \max(y_{i-1},T_i), y_i \leq D_i, a_i \in \mathcal{A}, \forall i \qquad \text{(ICK-CLO)}$$
$$\lim_{N\to\infty} \frac{1}{N}\sum_{i=1}^{N} \mathop{E}_{T_i,D_i,L_i,\mathcal{Q},C_i} w_i(x_i,y_i,a_i) \leq W$$

The optimization in the ICK-CLO problem is the same as the CK-CLO problem except that the ICK-CLO problem minimizes the expected average distortion for the infinite number of DUs over the expected average resource constraint. However, the solution to the ICK-CLO problem is quite different from the solution to the CK-CLO problem. In the following, we will first present the optimal solution to the ICK-CLO problem, and then we will compare this solution with that of the CK-CLO problem. Finally, we will develop an online cross-layer optimization for each DU.

*1) MDP formulation of the cross-layer optimization for infinite DUs*

Similar to the dual problem presented in Section III, the dual problem (referred to as ICK-DCLO) corresponding to the ICK-CLO problem is given by the following optimization.

$$\max_{\lambda \geq 0} g(\lambda), \qquad \text{(ICK-DCLO)}$$

where $g(\lambda)$ is computed by the following optimization.

$$g(\lambda) = \min_{x_i \geq \max(y_{i-1},T_i), y_i \leq D_i, a_i \in \mathcal{A}, \forall i} \lim_{N\to\infty} \frac{1}{N}\sum_{i=1}^{N} \mathop{E}_{\Psi_i, C_i}(Q_i(x_i,y_i,a_i) + \lambda w_i(x_i,y_i,a_i)) - \lambda W, \quad (14)$$

where the Lagrange multiplier $\lambda$ is associated with the expected average resource constraint, which is the same as the one in Eq. (1). Once the optimization in Eq. (14) is solved, the Lagrange multiplier is then updated as follows:

$$\lambda^{k+1} = \left\{\lambda^k + \alpha^k\left(\lim_{N\to\infty}\frac{1}{N}\sum_{i=1}^{N}\mathop{E}_{T_i,D_i,L_i,\mathcal{Q},C_i} w_i(x_i,y_i,a_i) - W\right)\right\}^+. \qquad (15)$$

Hence, in the following, we focus on the optimization in Eq. (14).

From the assumption presented at the beginning of Section V.A, we note that $T_{i+1} - T_i$, $D_i - T_i$, $C_i$ and other attribute of DU $i$ are i.i.d. random variables. Hence, for the independently decodable DUs, if we know the value of $T_i$, the attributes and network conditions of all the future DUs (including DU $i$) are independent of the attributes and network conditions of previous DUs. As shown in Figure 4, DU $i-1$ will impact the cross-layer action selection of DU $i$ only through ETX $y_{i-1}$ since $x_i = \max(y_{i-1}, t_i)$. In other words, DU $i-1$ brings forward or postpones the transmission of DU $i$ by determining its ETX $y_{i-1}$. If we define a state for DU $i$ as $s_i = \max(y_{i-1} - t_i, 0)$. Then, the impact from previous DUs is fully characterized by this state. Knowing the state $s_i$, the cross-layer optimization of DU $i$ is independent of the previous DUs. This observation motivates us to model the cross-layer optimization for the time-varying DUs as a MDP [13] in which the state transition from state $s_i$ to state $s_{i+1}$ is determined only by the ETX $y_i$ of DU $i$ and the time $t_{i+1}$ DU $i+1$ is ready for transmission, i.e. $s_{i+1} = \max(y_i - t_{i+1}, 0)$. The action in this MDP formulation is the STX $x_i$, ETX $y_i$ and the action $a_i$. The STX is automatically set $x_i = \max(y_{i-1}, t_i)$. The immediate cost by performing the cross-layer action is given by $Q_i(x_i, y_i, a_i) + \lambda w_i(x_i, y_i, a_i)$.

Given the resource price $\lambda$, the optimal policy (i.e. the optimal cross-layer action at each state) for the optimization in Eq. (14) satisfies the dynamic programming equation [13], which is given by

$$V(s) = \mathop{E}_{D,L,\mathcal{Q},C,T}\left\{\max_{\substack{x=s+t \\ y<D \\ a\in\mathcal{A}}}[Q(x,y,a) + \lambda w(x,y,a) + V(\max(y-T,0))]\right\} - \beta \quad (16)$$

where $V(s)$ represents state-value function at state $s$ and the difference $V(s) - V(0)$ represents the total impact that the previous DU impose on all the future DUs by delaying the transmission of the next DU by $s$ seconds; $t$ is the time the current DU is ready for transmission; and $\beta$ is the optimal average cost. It is easy to show that $V(s)$ is a non-decreasing function of $s$ because the larger the state $s$, the larger the delay in transmission of the future DUs, and therefore the larger the distortion.

There is a well-known relative value iteration algorithm (RVIA) [13] for solving the dynamic programming equation in Eq. (16), which is given by

$$V_{n+1}(s) = \mathop{E}_{D,\mathcal{Q},C,T}\left\{\max_{x=s+t, y<D, a\in\mathcal{A}}[Q(x,y,a) + \lambda w(x,y,a) + V_n(\max(y-T,0))]\right\} - V_n(0) \quad (17)$$

where $V_n(\cdot)$ is the state-value function obtained at the iteration $n$.

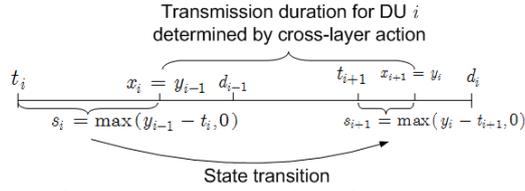

Figure 4. State of DU $i$ and state transition from DU $i$ to DU $i+1$

*2) Comparison of the solutions to CK-CLO and ICK-CLO*

In this section, we discuss the similarity and difference between the solutions to the CK-CLO and ICK-CLO problems. We note that both solutions are based on the duality theory and solve dual problems instead of the original constrained problems. Hence, both solutions use the resource price to control the amount of resource used for each DU.

In the CK-CLO problem, the solution is obtained assuming complete knowledge about the DUs' attributes and the experienced network conditions, which is not available for the ICK-CLO problem. Hence, in the DUCLO for the CK-CLO problem, the impact on the neighboring DUs is fully characterized by scalar numbers $\mu_{i-1}$ and $\mu_i$. The cross-layer action selection for each DU is based on the assumption that the cross-layer actions for neighboring DUs (previous and future DUs) are fixed. However, in the RVIA for the ICK-CLO problem, the cross-layer action selection for each DU is based on the assumption that the cross-layer actions for the previous DUs are fixed (i.e. the sate $s$ is fixed) and the future DUs (and the cross-layer actions for them) are unknown. The impact from the previous DUs is characterized by the state $s$ and the impact on the future DUs is characterized by the state value function $V(s)$.

Hence, the solution to the CK-CLO problem cannot be generalized to the online DUCLO which has no exact information about the future DUs. However, the solution to the ICK-CLO problem can be easily extended to the online cross-layer optimization for each DU, since it takes into account the stochastic information about the future DUs once it has the state value function $V(s)$. In the next section, we will focus on developing the learning algorithm for updating the state-value function $V(s)$.

*3) Online cross-layer optimization using learning*

In this section, we develop an online learning to update the state-value function $V(s)$ and the resource price $\lambda$. Assume that, for DU $i$, the estimated state-value function and resource price are denoted by $V_i(s)$ and $\lambda_i$, then the cross-layer optimization for DU $i+1$ is given by

$$\min_{x_i,y_i,a_i} Q_i(x_i,y_i,a_i) + \lambda_i w_i(x_i,y_i,a_i) + V_i(\max(y_i - t_{i+1}, 0))$$
$$s.t. \quad x_i = s_i + t_i, y_i \leq d_i, a_i \in \mathcal{A} \tag{18}$$

This optimization can be solved as in Section III.C. The remaining question is how we can choose the right price of resource $\lambda_i$ and estimate the state-value function $V_i(s)$.

From the theory of stochastic approximation [22], we know that the expectation in Eq. (17) can be removed and the state-value function can be updated as follows:

$$V_{i+1}(s_i) = (1 - \gamma_i)V_i(s_i) +$$
$$\gamma_i \left\{ \max_{x_i=s_i, y_i<d_i, a_i \in \mathcal{A}} [Q_i(x_i,y_i,a_i) + \lambda w_i(x_i,y_i,a_i) + V_i(\max(y_i - t_{i+1}, 0))] - V_i(0) \right\}, \tag{19}$$
$$\text{and } V_{i+1}(s) = V_i(s), \text{if } s \neq s_i$$

where $\gamma_i$ satisfies $\sum_{j=1}^{\infty} \gamma_j = \infty, \sum_{j=1}^{\infty} (\gamma_j)^2 < \infty$. We should note that, in this proposed learning algorithm, the cross-layer action of each DU is optimized based on the current estimated state-value function and resource price. Then the state-value function is updated based on the current optimized result. Hence, this learning algorithm does not explore the whole cross-layer action space like the Q-learning algorithm [26] and may only converge to the local solution. However, in the simulation section, we will show that it can achieve the similar performance as the CK-CLO with $M = 10$, which means that the proposed online learning algorithm can forecast the impact of current cross-layer action on the future DUs by updating the state-value function.

Since $V_i(s)$ is a function of the continuous state $s$, the formula in Eq. (19) cannot be used to update state-value function for each state. To overcome this obstacle, we use a function approximation method similar to the work in [19] to approximate the state-value function by a finite number of parameters. Then, instead of updating the state-value function at each state, we use the formula in Eq. (19) to update the finite parameters of the state-value function. Specifically, the state-value function $V(s)$ is approximated by a linear combination of the following set of feature functions:

$$V(s) \approx \begin{cases} \sum_{k=1}^{K} r^k v^k(s) & \text{if } s \geq 0 \\ 0 & \text{o.w.} \end{cases} \tag{20}$$

where $r = [r^1, \cdots, r^K]'$ is the parameter vector; $v(s) = [v^1(s), \cdots, v^K(s)]'$ is a vector function with each element being a scalar feature function of $s$ [19]; and $K$ is the number of feature functions used to represent the impact function. The feature functions should be linearly independent. In general, the state-value function $V(s)$ may not be in the space spanned by these feature functions. The larger the value $K$, the more accurate this approximation. However, the large $K$ requires more memory to store the parameter vector. Considering that the state-value function $V(s)$ is non-decreasing, we choose $v(s) = \left[s^1, \cdots, \frac{s^K}{K!}\right]'$ as the feature functions. Using these feature functions, the parameter vector $r = [r^1, \cdots, r^K]'$ is then updated as follows:

$$\begin{aligned} r_{i+1} = (1-\gamma_i) r_i + \\ \gamma_i \left\{ \max_{x_i = s_i, y_i < d_i, a_i \in \mathcal{A}} [Q_i(x_i, y_i, a_i) + \lambda w_i(x_i, y_i, a_i) + V_i(\max(y_i - t_{i+1}, 0))] - V_i(0) \right\} v(s_i) \end{aligned} \tag{21}$$

Similar to the price update in Section III, the online update for $\lambda$ is given as follows:

$$\lambda_{i+1} = \left( \lambda_i + \kappa_i \left( \frac{1}{i} \sum_{j=1}^{i} w_j - W \right) \right)^+, \tag{22}$$

where $\kappa_i$ satisfies $\sum_{j=1}^{\infty} \kappa_j = \infty, \sum_{j=1}^{\infty} (\kappa_j)^2 < \infty, \lim_{j \to \infty} \frac{\kappa_j}{\gamma_j} = 0$.

In Eqs. (21) and (22), iterating on the state-value function $V(y)$ and the resource price $\lambda$ at different timescales ensures that the update rates of the state-value function and resource price are different. The resource price is updated on a slower timescale (lower update rate) than the state-value function. This means that, from the perspective of the resource price, the state-value function $V(y)$ appears to converge to the optimal value corresponding to the current resource price. On the other hand, from the perspective of the state-value function, the resource price appears to be almost constant.

The algorithm for the proposed online optimization using learning is illustrated in Algorithm 2.

Algorithm 2: Proposed online optimization using learning

**Initialize** $\lambda_1, r_1 = 0$, $s_1 = 0$, $i = 1$
**For** each DU $i$
   Observe the attributes and network condition of DU $i$ and the time $t_{i+1}$ at which DU $i+1$ is ready for transmission;
   Layered solution to the DUCLO given in Eq. (18);
   Update $s_{i+1} = \max(y_i - t_{i+1}, 0)$, $\lambda_{i+1}$ as in Eq. (22) and $r_{i+1}$ as in Eq. (21);
   $i \leftarrow i + 1$
**End**

*B. Online optimization for interdependent DUs*

In this section, we consider the online cross-layer optimization for the interdependent DUs as discussed in Section IV. In order to take into account the dependencies between DUs, we assume that the DAG of all DUs is known a priori. This assumption is reasonable since, for instance, the GOP structure in video streaming is often fixed. When optimizing the cross-layer action $(x_i, y_i, a_i)$ of DU $i$, the transmission results $p_k(x_k^*, y_k^*, a_k^*)$ and $e_k(x_k^*, y_k^*, a_k^*)$ of DUs with index $k < i$ are known. Then, the sensitivity $Q_i'(x_i, y_i, a_i)$ of DU $i$ is computed, based on the current knowledge, as follows:

$$Q_i'(x_i, y_i, a_i) = q_i p_i(x_i, y_i, a_i) \prod_{k \prec i} \left(1 - e_k(x_k^*, y_k^*, a_k^*)\right) - (1 - e_i(x_i, y_i, a_i)) \left\{ \sum_{i' \succ i} \tilde{q}_{i'}(1 - \tilde{p}_{i'}) \prod_{\substack{j \prec i' \\ j \neq i}} \left(1 - \tilde{e}_j(x_j, y_j, a_j)\right) \right\}, \quad (23)$$

where $\tilde{q}_{i'}(1 - \tilde{p}_{i'})$ is the estimated distortion impact of DU $i'$. The term $e_k(x_k^*, y_k^*, a_k^*)$ is the error propagation function of DU $k < i$, which is already known. If $j < i$, $\tilde{e}_j(x_j, y_j, a_j) = e_j(x_j^*, y_j^*, a_j^*)$, otherwise $\tilde{e}_j(x_j, y_j, a_j) = 0$ by assuming that DU $j$ can be successfully received. In other words, if DU $k$ is transmitted, the transmitted results $p_k(x_k^*, y_k^*, a_k^*)$ and $e_k(x_k^*, y_k^*, a_k^*)$ are used, otherwise DU $k$ is assumed to be successfully received in the future.

Similar to the online cross-layer optimization for independent DUs given in Section V.A, the online optimization for the interdependent DUs is given as follows:

$$\begin{aligned} \min_{x_i, y_i, a_i} \quad & Q_i'(x_i, y_i, a_i) + \lambda w_i(x_i, y_i, a_i) + V_i(\max(y_i - t_{i+1}, 0)) \\ s.t. \quad & x_i = s_i + t_i, y_i \leq d_i, a_i \in \mathcal{A} \end{aligned} \quad (24)$$

The update of the parameter vector $r$ and the resource price $\lambda$ is the same as in Eqs. (21) and (22).

## VI. NUMERICAL RESULTS

In this section, we present our numerical results to evaluate the proposed decomposition method and the online algorithm. We consider an example in which the user streams the delay sensitive DUs over a time-varying channel with energy constraints.

### A. *Models for distortion impact and energy cost functions*

In this example, we consider the proposed cross-layer optimization solution to determine the optimal scheduling and energy allocation for DUs with various attributes at the application layer transmitted over a time-varying channel at the physical layer. The transmission action is the number of bits, $a_i$, to be transmitted. The consumed energy (cost) is given, as in [5], by

$$w_i(x_i, y_i, a_i) = \frac{N_0}{c_i}\left(2^{\frac{a_i}{y_i - x_i}} - 1\right)(y_i - x_i), \tag{25}$$

where $N_0$ denotes thermal noise. It is easy to show that $w_i(x_i, y_i, a_i)$ is a convex function of the difference $y_i - x_i$ and $a_i$.

We assume that the application data is compressed in a scalable way [11] such that, given the amount of transmitted bits, $a_i$, the expected distortion of the independent DU with index $i$ is given, as in [18], by

$$Q_i(x_i, y_i, a_i) = q_i 2^{-\theta_i \min(a_i, l_i)}, \tag{26}$$

where $\theta_i > 0$. That is, $p_i(x_i, y_i, a_i) = 2^{-\theta_i \min(a_i, l_i)}$. It is easy to show that $p_i(x_i, y_i, a_i)$ is a convex function of $a_i$.

For interdependent DUs, the expected distortion of DU $i$ is then given by

$$Q_i(x_i, y_i, a_i) = q_i\left[1 - \prod_{k \preceq i}\left(1 - 2^{-\theta_k \min(a_k, l_k)}\right)\right] \tag{27}$$

That is, $e_k(x_k, y_k, a_k) = 2^{-\theta_i \min(a_i, l_i)}$ [8]. The distortion reduction for each DU is given by $q_i - Q_i$.

In this example, the distortion impact $q_i$ is the realization of a uniformly distributed random variable in the range of $[50, 150]$. The DU size $l_i$ is assumed to be constant and equals 10000bits. The varying DU size is considered in Section VI.F for video streaming. The arrival interval $t_i - t_{i-1}$ is the realization of an exponentially distributed random variable with the mean of 50 ms. The DU lifetime $d_i - t_i$ is 50 ms. The parameter $\theta_i$ equals 0.5. We will verify the efficiency of the proposed methods using the model developed in this section in Sections B~ E. We will further consider a more realistic scenario with video streaming over wireless networks in Section F.

---

[8] Here the error propagation function represents the fact that increasing the faction of DU $i$ reduces the amount of error propagated to other DUs.

B.  *Dual and primal solutions and duality gap for independent DUs*

Figure 5 (a) shows the duality gap between the dual solutions and primal solutions over 110 iterations in a setting with $M = 10$ independent DUs. It is shown that the duality gap goes to zero after around 100 iterations, which demonstrates that the subgradient algorithm developed in Section III converges to the optimal total expected distortion given by the primal solutions. Figure 5 (b) further shows that the primal and dual solutions are equivalent. However, the subgradient method requires around 100 iterations to converge to the optimal solutions, which may be hard to implement in the real-time applications (e.g. video streaming) since it requires a lot of computation. Hence, in Section V, we have developed an online algorithm which can significantly reduce the complexity of the cross-layer optimization (i.e. one iteration) and only use the current available information. The simulation results for the online algorithms are presented in Section VI.D.

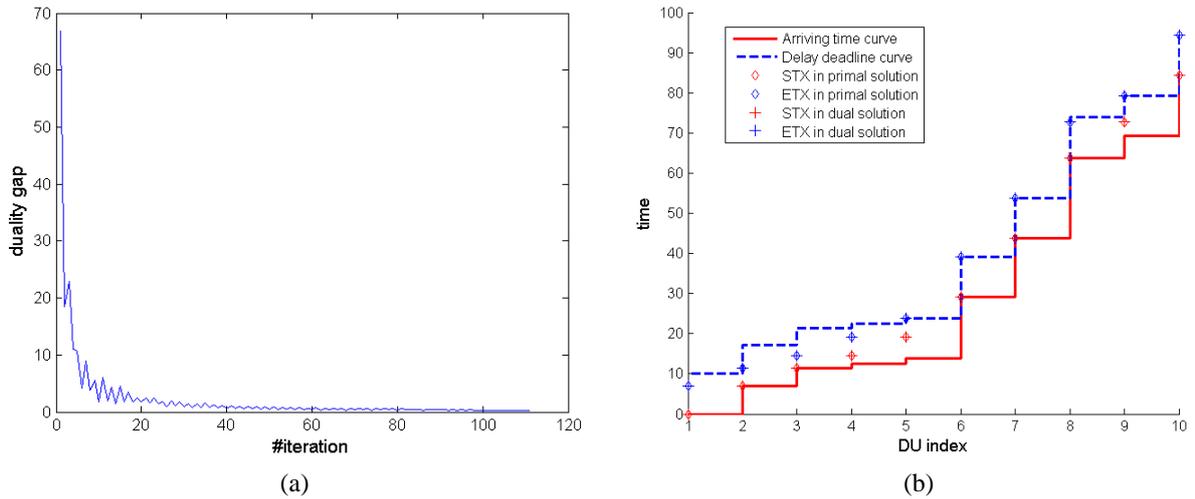

Figure 5.  (a) Duality gap between the dual and primal solutions for independent DUs; (b) Dual and primal optimal scheduling time for independent DUs

*C. Dual and primal solutions and duality gap for the interdependent DUs*

Figure 6 (a) shows the duality gap between the dual solutions and primal solutions for the interdependent DUs with $M = 10$. Although the cross-layer optimization problem for the interdependent DUs is not a convex optimization, it is shown here that the duality gap in this example goes to zero after around 230 iterations, which demonstrates that the subgradient algorithm developed in Section III also converges in the cross-layer optimization for interdependent DUs. The subgradient algorithm for the interdependent DUs requires two types of iterations: one is the outer iteration which updates the price of the resource $\lambda$ and NIFs $\mu$ and the other one is the inner iteration which is to find the optimal cross-layer action for each DU given $\lambda$ and $\mu$ as shown in Eq. (13). Figure 6 (b) shows the required number of inner iterations per outer iteration using the cross-layer actions obtained in the previous outer iteration as the starting point in the current outer iteration. It is clear that 2~6 inner iterations are required for each outer iteration to converge to the optimal cross-layer actions given $\lambda$ and $\mu$. Hence, the subgradient method requires a total of 651 inner iterations, which is unacceptable for the real-time applications (e.g. video streaming). As discussed in Section VI.B, this motivates us to develop an online algorithm which was presented in Section V. The simulation results for the online algorithm are presented in Section VI.E.

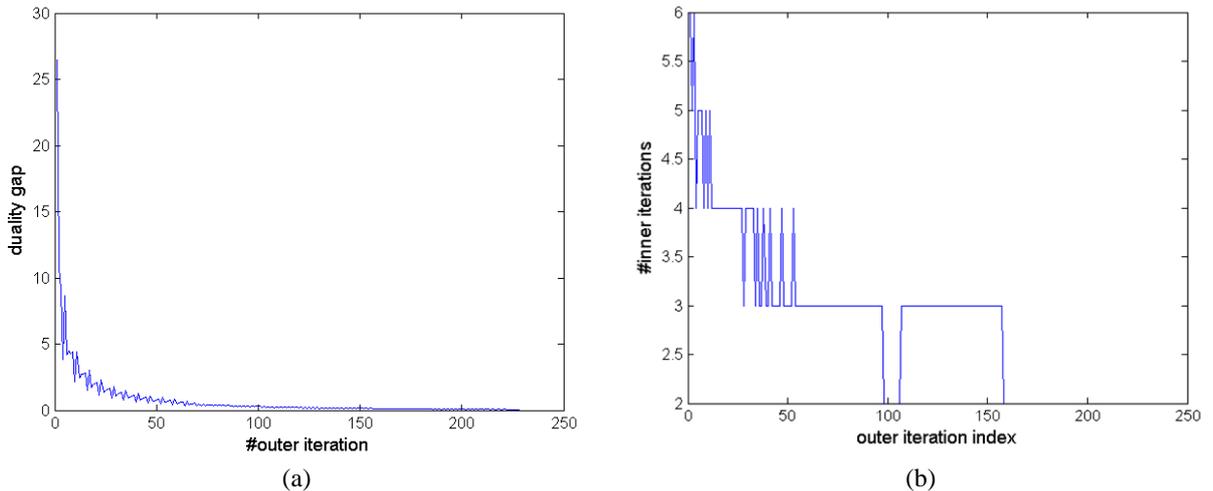

(a) (b)
Figure 6. (a) Duality gap between the dual and primal solutions for interdependent DUs, (b) Number of inner iterations per outer iterations for the cross-layer optimization of interdependent DUs

*D. Online cross-layer optimization for independent DUs*

In this simulation, we consider three online algorithms for the scenario with independent DUs. The first is the online cross-layer optimization for each DU proposed in Section V. The second performs the cross-layer optimization every $M=10$ DUs by assuming complete knowledge of these $M$ DUs' attributes and underlying network conditions (we call this $M$-DU cross-layer optimization). The third one performs the cross-layer optimization for each DU (i.e. $M=1$, called myopic online optimization). We will refer to the transmission of 10 DUs as one cycle.

Figure 7 depicts the distortion reduction of each cycle under various resource constraints for these three algorithms. From this figure, we note that, on the one hand, the online cross-layer optimization proposed in Section V outperforms the myopic online optimization by around 6% for various energy constraints because the proposed online optimization can predict the impact on the future DUs through the state-value function and allocate the energy for each cycle based on the importance of DUs. On the other hand, the $M$-DU cross-layer optimization outperforms the proposed online cross-layer optimization by around 4% since $M$-DU cross-layer optimization explicitly considers the exact information of future DUs which is not available in the online cross-layer optimization. However, the proposed online cross-layer optimization has the following advantages, compared to the $M$ DU cross-layer optimization: (i) it performs the cross-layer optimization for each DU and updates $\lambda$ and state-value function $V(s)$ for each DU without requiring multiple iterations, which significantly reduces the computational complexity; (ii) it does not require exact information about the future DUs' attributes and network conditions.

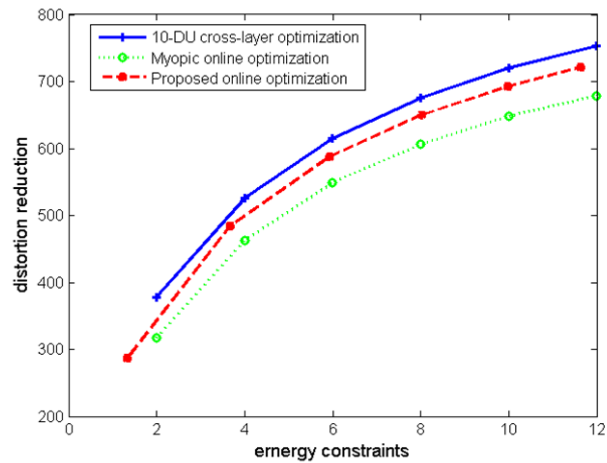

Figure 7. The distortion reduction under various energy constraints for independent DUs

*E. Online cross-layer optimization for interdependent DUs*

In this simulation, we also consider three online algorithms as described in Section VI.D for the scenario with interdependent DUs. The interdependencies (represented by a DAG) are generated randomly every 10 DUs. The interdependency between DUs happens only within one cycle (for instance, a cycle could represent one group of pictures (GOP) of the video sequences). Figure 8 shows the distortion reduction of each cycle under various energy constraints. From this figure, we note that, for interdependent DUs, our proposed online cross-layer optimization can significantly improve the performance (more than 28% increased) compared to the myopic online optimization, and has similar performance as the $M$-DU optimization. We further show the distortion reduction and energy allocation for each cycle when the average energy constraint is 10 (i.e. $W = 10$) in Figure 9. From this figure, we observe that, after the initial learning stage (about 30 cycles), our proposed online solution achieves the similar performance as the $M$-DU solution. We will also verify this observation in a more realistic scenario which is presented in the next section. The reason that our proposed solution can have similar performance as the $M$-DU solution is as follows: for the interdependent DUs, the amount of the distortion reduction is mainly determined by the important DUs (on which many other DUs depend on) and our solution can ensure that more important DUs are successfully transmitted by allocating more energy to them.

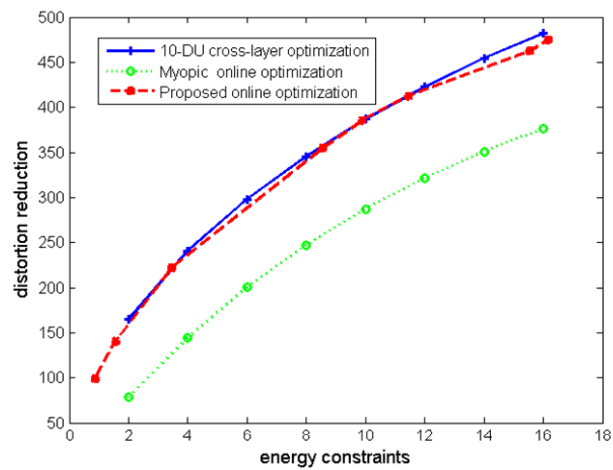

Figure 8. Distortion reduction under various energy constraint for interdependent DUs

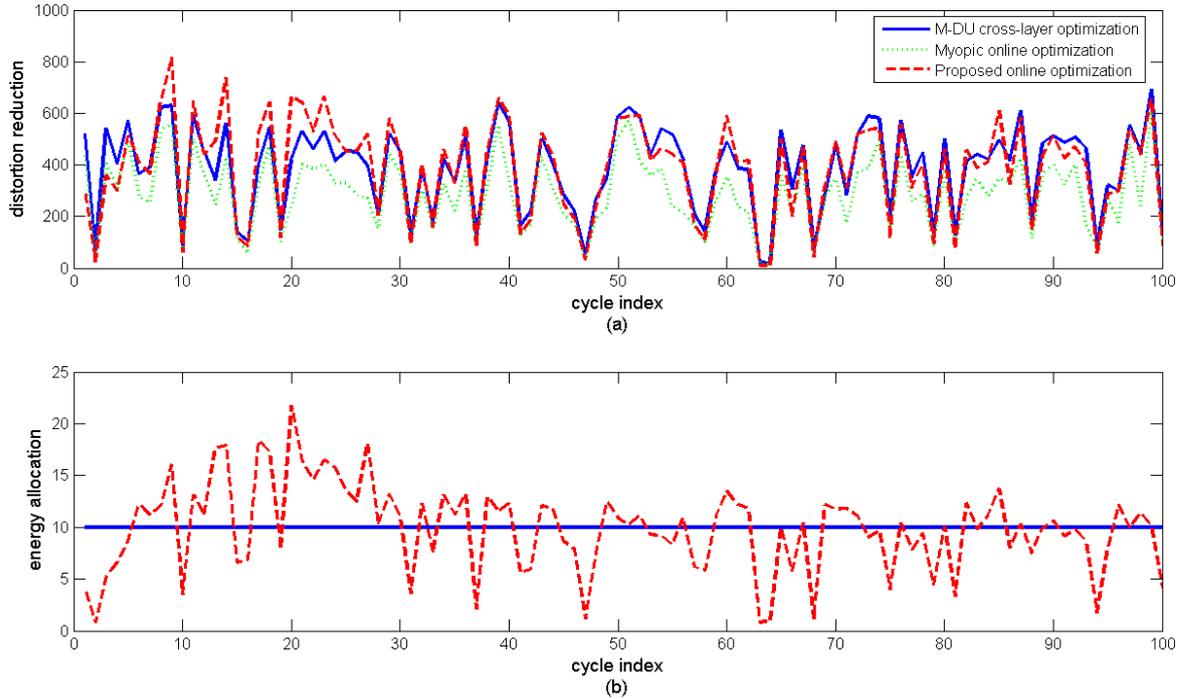

Figure 9. (a) Distortion reduction and (b) average energy consumption for each cycle.

*F. Online cross-layer optimization for video streaming*

In this simulation, we consider a more realistic situation in which the wireless user streams the video sequence "Coastguard" (CIF resolution, 30 Hz) over the time-varying wireless channel. For the compression of the video sequence, we used a scalable video coding schemes based on Motion Compensated Temporal Filtering (MCTF) using wavelets [25]. Such 3D wavelet video compression is attractive for wireless streaming applications because it provides on-the-fly adaptation to channel conditions, support for a variety of wireless receivers with different resource capabilities and power constraints, and easy prioritization of various coding layers and video packets. We consider every 8 frames as one GOP and each DU corresponds to one frame at a certain temporal level, as shown in [11]. The dependency between DUs is illustrated in Figure 10 (a). We compare three online optimization methods as in Section VI.E. Figure 10 (b) depicts the received Peak Signal-to-Noise Ratio (PSNR) in dB under these methods. From this figure, we note that the myopic online optimization achieves the PSNR of 27.1dB on average which is generally considered very poor video quality. However, our proposed online cross-layer optimization can improve the video quality over time through the learning procedure and achieve the PSNR of 29.9 dB (2.8dB better than the myopic solution[9]). Moreover, the achieved video quality in our solution is much smoother (i.e. the PSNRs of all the frames do not vary dramatically like in the myopic case). We also demonstrated that the proposed solution achieves the similar performance (only 0.5dB less on average) as the $M$-DU method, as indicated in Section VI.E.

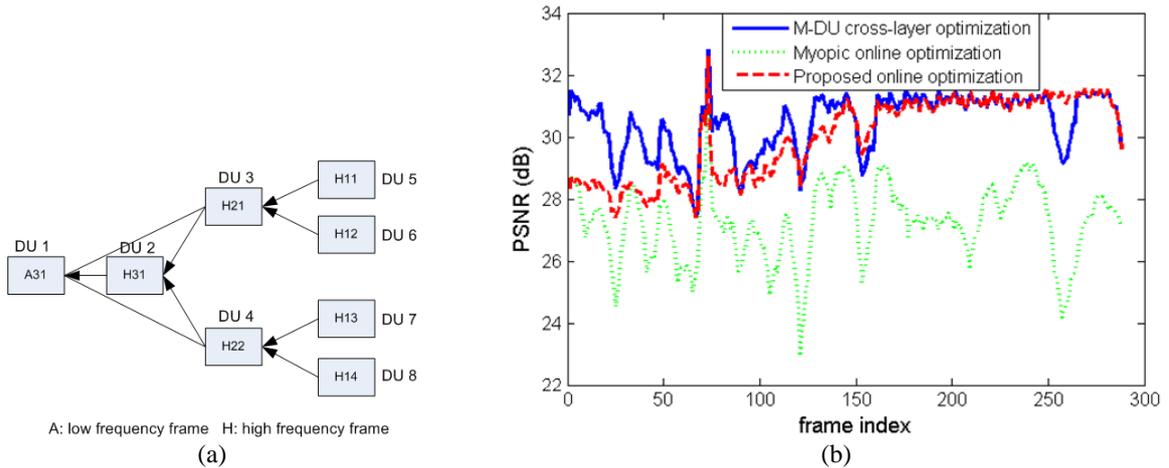

(a)            (b)
Figure 10. (a) DAG for the interdependency between DUs with one GOP; (b) PSNR for the video sequence "coastguard" under three cross-layer optimization methods

## VII. CONCLUSIONS

In this paper, we consider the problem of cross-layer optimization for delay-sensitive applications, and we develop decomposition principles that guarantee the optimal performance of the application while requiring the necessary message exchanges between neighboring DUs and between layers. To account for the unknown and dynamic characteristics of real-time delay-sensitive applications, we further propose an efficient online cross-layer optimization with low complexity, which can be used for live events (e.g. real-time encoding and streaming of ongoing events, videoconferencing etc.), when the encoding is done in real-time and the wireless user does not have a priori information about future application data and network conditions.

---

[9] Note that it is well known that 0.5 dB performance improvement is visible for a trained observer, 1dB performance improvement is visible for any observer and 2dB of more results in significantly visible performance improvements.